\documentstyle[epsfig]{mn}

\title[Quasar Enrichment and First Stars]{Early Enrichment of Quasars by First Stars}

\author[A. Venkatesan, R. Schneider \& A. Ferrara]{
A. Venkatesan$^{1,2}$, R. Schneider$^{3,4}$, A. Ferrara$^{5}$ \\
$^1$ CASA, Department of Astrophysical and Planetary Sciences,
 University of Colorado, UCB 389, Boulder, CO 80309-0389 \\
$^2$ NSF Astronomy and Astrophysics Postdoctoral Fellow \\
$^3$ INAF/Osservatorio Astrofisico di Arcetri, Largo E. Fermi 5, 50125
  Firenze, Italy \\
$^4$ Centro Enrico Fermi, Via Panisperna 89/A, 00184 Roma, Italy \\
$^5$ SISSA/International School for Advanced Studies, Via
  Beirut 4, 34014 Trieste, Italy \\}

\pagerange{\pageref{firstpage}--\pageref{lastpage}}
\pubyear{2003}
\begin{document}

\maketitle
\label{firstpage}

\begin{abstract}

Studies of the broad emission-line regions (BLRs) in quasars have
revealed solar or higher enrichment levels up to the highest
redshifts. In combination with the presence of large amounts of dust
in QSOs at $z \sim$ 6, this implies that substantial amounts of star
formation and nucleosynthesis took place at significantly earlier
epochs. Here, we examine whether a top-heavy stellar initial mass
function (IMF) is indicated by current data, by modelling the
contributions from different regions of the IMF, including Type Ia/II
and pair instability supernovae, to the metal synthesis in BLRs. We
find that, in order to reproduce the observations of roughly solar
values of N/C and Fe/Mg in these objects, (i) stars with a present-day
IMF are sufficient, regardless of their metallicity, (ii)
zero-metallicity stars with a top-heavy IMF severely underproduce N/C,
and (iii) the contribution of Type Ia SNe is not strongly required by
the data.  Therefore, stars of mass $\sim$ 1-40 $M_\odot$ must have
existed at $z \sim$ 10--20, possibly coeval with any hypothesized
stars of masses $\ga$ 100 $M_\odot$ at these epochs. This is in
agreement with the nucleosynthetic abundance pattern detected in
extremely metal-poor stars in the galactic halo.

\end{abstract}

\begin{keywords}
cosmology: theory - first stars - galaxies:
  abundances - galaxies: starburst - quasars: emission lines
\end{keywords} 

\section{Introduction}

Multiwavelength observations of increasing quality over the last decade
have begun to strongly constrain the link between the earliest epochs of
star formation in the universe and the metal enrichment of high-redshift
($z$) protogalaxies and the intergalactic medium (IGM). Constraints on
abundances at the highest redshifts currently comes from data on QSOs,
where the chemical composition of the gas can be estimated from spectral
features from the broad emission-line regions (BLRs) as well as broad and
narrow absorption lines, probing regions ranging from sub-pc scales to
those exceeding several tens of parsecs.

Considerable progress has been made in recent years in observing and
modelling such QSO environments, and detailed chemical evolution models
have been developed to better assess reliable abundance diagnostics (Hamann
\& Ferland 1999).  The NV/CIV line ratio has received particular attention
because of its high sensitivity to the overall metallicity of the gas. For
typical values of BLR parameters, the models predict NV/CIV $\sim$ 0.1 for
solar N/C. The current data indicates that most QSOs have supersolar N/C
and BLR gas metallicities, including the highest-$z$ ones at $z \sim$ 6.28
and 5.99, which have NV/CIV = 0.35 and 0.67 respectively (Pentericci et
al. 2002).  In addition, a roughly constant value of FeII/MgII ($\sim$
2--15) is detected in QSOs up to the highest redshifts at $z \sim$ 6.4
(Freudling, Corbin \& Korista 2003; Maiolino et al. 2003; Dietrich et
al. 2003a and references therein). Assuming that FeII/MgII traces Fe/Mg and
using FeII/MgII $\sim$ 3 for solar Fe/Mg, the data imply BLR gas
metallicities of near- or exceeding solar values (Dietrich et al. 2003b).
Finally, recent submm/mm/IR observations of high-z QSOs have revealed the
presence of large amounts of dust (thermal dust masses $\sim 10^8 M_\odot$)
up to redshifts as high as 6.4 (Bertoldi et al. 2003), implying associated
star formation rates of $\sim 10^3 M_\odot$ yr$^{-1}$.

Although the sizes and masses of the physical regions probed by BLR and
dust observations are orders of magnitude apart, together they yield a
consistent scenario: that significant stellar activity and nucleosynthesis
occurred at $z \ga$ 6. At these epochs, the age of the universe is
approximately 1 Gyr and approaches the enrichment timescale of low-mass
stars and Type Ia supernovae (SNe). These data may imply that star
formation in such environments occurred rapidly and involved an initial
mass function (IMF) favoring massive stars (Hamann \& Ferland 1993).  Our
aim here is to test whether the observed abundance ratios in combination
with the constraints from stellar evolutionary timescales can place
interesting limits on the nature of the stellar IMF at early epochs.  In
particular, the presence of significant quantities of iron has led some
authors to propose a Type Ia SN origin. Alternatively, Type II SNe and pair
instability SNe (PISNe; Woosley \& Weaver 1995; Heger \& Woosley 2002) from
stars of respective mass ranges 10--40 M$_\odot$ and 140--160$M_\odot$
could also be substantial sources of iron in QSO BLRs. These massive stars
have the added advantages of providing the same nucleosynthesis site for
alpha-elements such as magnesium, and of being sources that can create dust
and generate prompt metal enrichment.

In this work, we examine the evolution of N/C and Fe/Mg for cases involving
a top-heavy and present-day IMF in a burst scenario at early epochs. Our
approach here is simple and chosen to reveal whether a preferred stellar
mass range is indicated by the available data. For more detailed modelling
of various star formation scenarios, we refer the reader to, e.g., Romano
et al. (2002), Hamann \& Ferland (1999) and references therein.

\section{Stellar IMF Models}

We take the stellar IMF to have the form, $\phi(M) = \phi_0 M^{-\alpha}$,
where $\alpha$ and $M$ are the IMF slope and stellar mass respectively. We
consider two cases: (1) a present-day form, where $\alpha$ equals the
Salpeter value of 2.35, and the stellar mass range is 1--100 $M_\odot$, and
(2) a flat top-heavy IMF involving very massive stars (VMSs) with $\alpha$
= 1 (ensuring a flat mass-weighted IMF, $M \phi(M)$) in the mass range
100--1000 $M_\odot$. We have chosen this division in order to distinguish
the stellar signatures of these two IMFs, although it is possible that
these IMFs were coeval in the past, given the right combination of
conditions (Scannapieco, Schneider \& Ferrara 2003). Both cases are
normalized over their respective mass ranges as, $ \int dM M \phi(M) = 1
$. For the first case, we consider two stellar metallicities, $Z_\star$ = 0
and $Z_\star = Z_\odot$, and only $Z_\star$ = 0 for the second, since it is
currently thought that stars preferentially form in a top-heavy IMF with
characteristic masses of a few 100 $M_\odot$ when the gas metallicity is
below a critical value of $Z_{\rm cr} = 10^{-5 \pm 1} Z_\odot$ (Bromm et
al. 2001; Schneider et al. 2002, 2003a).

%%%%%%%%%%%%%%%%%%%%%%%%%
\begin{table}
\begin{center}
\caption{
Properties of the three stellar IMF models considered here: $M_{\rm
low}$ and $M_{\rm high}$ are the lower and upper mass limits in $M_{\odot}$
of the stellar mass range, $\alpha$ is the exponent of the power-law IMF
(see text), $Z_\star$ is the initial stellar metallicity and the last entry
indicates whether the contribution from Type Ia SNe is considered.}
\begin{tabular}{|c|cccc|c|}\hline 
Model & $M_{\rm low}$ & $M_{\rm high}$ & $\alpha$ & $Z_\star$ & Type Ia \\ \hline

  A   &  1        & 100      &  2.35    & 0         &  no     \\
  B   &  1        & 100      &  2.35    & $Z_{\odot}$ &  yes    \\
  C   &  100      & 1000     &   1      & 0         &  no     \\
\hline 
\end{tabular}
\end{center}
\end{table}
%%%%%%%%%%%%%%%%%%%%%%%%%%%%%%%%%%%%%%%%%%%%%%%

The adopted ejected masses in C, N, Mg and Fe as a function of stellar
mass and metallicity are collected in Table~2.  We follow Venkatesan
\& Truran (2003) and use van den Hoek \& Groenewegen (1997) for
stellar masses $<$ 8 M$_\odot$. Since no detailed compilation exists
for $Z_\star = 0$ intermediate mass star (IMS) yields, we approximate
this by using $Z_\star=0.001$ (0.05 $Z_\odot$) IMS yields for
$Z_\star=0$, which do not appear to be substantially different by mass
fraction (see Venkatesan \& Truran 2003 and Chieffi et al. 2002). The
yields from Type II SNe for $Z_\star = Z_\odot$ ($Z_\star$ = 0) stars
in the 11(12)--40 M$_\odot$ mass range are taken from the case A
models from Woosley \& Weaver (1995), corresponding to typical SN
explosion energies of $\sim 1.2 \times 10^{51}$ erg. We assume that
stars of mass 40--100 $M_\odot$ implode directly into black holes and
therefore neglect their metal yields. For our top-heavy IMF case,
only the stars of mass 140--260 $M_\odot$ avoid complete collapse into
a black hole (Heger \& Woosley 2002) and contribute to the
nucleosynthetic output. We note that our adopted yields do not include
the effects of stellar rotation which could potentially be important,
particularly at low metallicities (Meynet \& Maeder 2002; Chiappini,
Matteucci and Meynet 2003). Our intent here is to consistently compare
the yields as a function of stellar mass and metallicity.

For Type Ia SNe, we use the revised W7 models from Nomoto et
al. (1997), which are essentially independent of the progenitor
mass. We assume that 5\% of the stellar IMF in the mass range 3--16
$M_\odot$ is in binary systems which lead to Type Ia SNe through the
single degenerate scenario (Gibson 1997; Matteucci \& Recchi 2001),
and that the mass ratio distribution of the binary's stars is close to
unity, with the metal input from Type Ia SNe occurring on the
evolutionary timescales of the secondary star. Recent theoretical
studies (e.g., Kobayashi et al. 1998) indicate that low-$Z$ ([Fe/H]
$\la$ -1) progenitors experience an inhibition of the Type Ia SN
phenomenon. Thus, we assume that Type Ia SNe do not occur for
$Z_\star=0$ stars. In combination with the two IMF cases above, this
leads to the three models (A, B and C) that we consider in this paper,
as detailed in Table 1. Although there remains significant uncertainty
in the exact values of stellar yields as a function of mass and
metallicity, these cases should roughly bound the contributions from
various stellar sites of origin.

For the elements considered here, we include the total abundances of
all the relevant isotopes of each element in its final yield; for
iron, this additionally includes $^{56}$Ni and $^{57}$Ni which
eventually decay into iron isotopes. Although the isotopes of most
elements are rare compared to the dominant form, some exceptions of
interest here are N, Mg and Fe.  We additionally note that the term
yield in this paper connotes the total ejected mass fraction in
individual elements or in metals; in the case of IMSs, since the
yields from van den Hoek \& Groenewegen (1997) correspond to the net
metal output, we account for the original metal composition of the
star as well.

Other than the input from Type Ia SNe which is modelled as described
above, the timescales for stellar yields correspond directly to the
main-sequence stellar lifetimes for a burst mode of star
formation. That is, for a single burst at $z$ = 30, the yield at $z$ =
6 equals the cumulative yield from those stars whose lifetimes do not
exceed the time difference between $z$ = 30 and $z$ = 6. The
main-sequence lifetimes are taken from Schaller et al. (1992) for
$Z_\star = Z_\odot$ stars and from Schaerer (2002) for $Z=0$ stars.

%%%%%%%%%%%%%%%%%%%%%%%%%%%%%%%%%%%%
\begin{figure}
\center{{\epsfig{figure=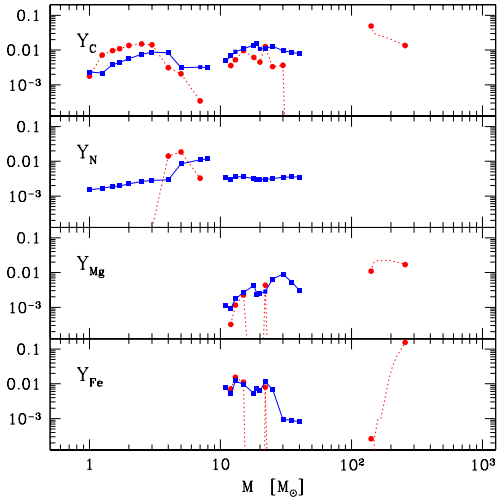,height=7.5cm}}}
\caption{Ejected mass fractions $Y_i=M_i/M$ as a function of stellar
mass, $M$, for two values of $Z_\star$, where $i$ can be C, N, Mg and Fe.
Dashed lines (circle points) and solid lines (square points) represent
$Y_i$ for $Z_\star=0$ and $Z_\star = Z_\odot$ stars, respectively.}
\end{figure}
%%%%%%%%%%%%%%%%%%%%%%%%%%%%%%%%%%%%

\section{Results}

Figure 1 displays the ejected mass fractions $Y_i=M_i/M$ as a function of
$M$ for two values of $Z_\star$ (0 and $Z_\odot$), where $i$ can be C, N,
Mg and Fe. We see that carbon has contributions from the entire IMF,
whereas the production of N is primarily limited to the part of the IMS
range ($\ga$ 3--4 $M_\odot$) that corresponds to efficient horizontal
branch burning (Henry 2003).  Mg and Fe production, not counting Type Ia
SNe, is limited to stellar masses $\ga$ 10 $M_\odot$: Type II SNe eject
comparable mass fractions in Mg and Fe, whereas PISNe produce an almost
mass-independent $Y_{\rm Mg}$ and a $Y_{\rm Fe}$ that rapidly increases
with stellar mass.  Figure 1 also shows the well-known trend that metal
production increases with decreasing $Z_\star$ for IMSs (Henry 2003) but
rises with increasing $Z_\star$ for massive stars (Woosley \& Weaver 1995).
For comparison, the yields available for $Z_\star = 0$ IMSs (Chieffi et
al. 2002) for Reimers mass loss parameter $\eta$ = 3 are $Y_{\rm C}$ =
[5.3,3] $\times 10^{-3}$ and $Y_{\rm N}$ = [3,4] $\times 10^{-3}$ for
stellar masses 4 and 7 $M_\odot$ respectively.

In order to facilitate direct comparison with observations, Figures 2
and 3 show the evolution of the element ratios Fe/Mg and N/C by mass
as a function of redshift and cosmological age for the three models A,
B and C specified in section 2.  Also shown in each figure is a shaded
rectangle indicating the approximate observed range of BLR gas
metallicities from studies of each abundance ratio, $\sim$ 1--10
$Z_\odot$ from NV/CIV over $z \sim$ 0.0--6.3 (Hamann \& Ferland 1999;
Pentericci et al. 2002), and $\sim$ 1--5 $Z_\odot$ from FeII/MgII over
$z \sim$ 0.1--6.4 (Freudling et al. 2003; Maiolino et al. 2003). The
solar values for N, Fe, and Mg were taken from the revised
photospheric abundances in Holweger (2001) and that for C from Allende
Prieto, Lambert \& Asplund (2002), so that (Fe/Mg)$_\odot$ $\sim$ 1.9
and (N/C)$_\odot$ $\sim$ 0.4 by mass. The rectangle in the plots is
intended to provide a rough reference for comparison between the
models here and the current data indications of $\sim$ solar or higher
values of N/C and Fe/Mg in QSO BLRs.  We emphasize that this box is
not an average of the data, which show significant scatter at a given
epoch and which measure the ratios of the observed fluxes in specific
element ionization states. Converting these, e.g., FeII/MgII, to the
underlying abundance ratios (like Fe/Mg) involves the modelling of
complex BLR environment physics (Hamann \& Ferland 1999) and related
uncertainties, which we have not accounted for here.

For models A and B, each set of curves for a specific $Z_\star$ corresponds
to the burst mode of star formation for three burst turn-on redshifts of
30, 20 and 10. These are chosen to be consistent with the range of
formation epochs of first stars from the $WMAP$ data on the microwave
background (Spergel et al. 2003) and from cosmological simulations (Ciardi,
Ferrara \& White 2003).  We follow the evolution for as long as it takes
all the stars in each IMF to evolve after the burst.  All the curves in
Figure 2, and that corresponding to model C in Figure 3, are artificially
extended to $z=0$ for comparison at $z \la$ 6.  The yields from VMSs are
assumed to be instantaneously generated, given the average VMS lifetime of
$\sim$ 2 Myr ($\Delta z \sim$ 0.4 at $z \sim$ 30).

Figures 2 and 3 show that massive stars and VMSs contribute trace
amounts of elements like N at high redshifts. As noted above, N, in
contrast to C, is primarily made from a narrower mass range in
metal-poor or metal-free stars, about 4--7 $M_\odot$. In Figure 3, we
see that for models A and B, after some initial fluctuations owing to
the detailed yields at the upper end of the IMF, N production peaks
when the burst age is about $10^8$ yr, corresponding to the lifetime
of a 4 $M_\odot$ star, the longest timescale on which N is created in
this case. C production, on the other hand, continues to rise
steadily, leading to the eventual steady decline of N/C. For Fe/Mg in
Figure 2, the situation is simpler for models A and C where the prompt
element generation from $\ga$ 10 $M_\odot$ stars is seen. The
curves for model B in Figure 2 display the increasing contribution of
Type Ia SNe at late times, peaking at cosmological ages corresponding
to the time lag for the evolutionary timescale of a $Z=Z_\odot$ 1.5
$M_\odot$ star, the smallest mass for the secondary star in the Type
Ia SN model that we consider here.

%%%%%%%%%%%%%%%%%%%%%%%%%%%%%%%%%%%%%%%%%%%%%%%%%%%%%%%%%%%%%%%%%%%%%%%%%%
\begin{figure}
\center{{\epsfig{figure=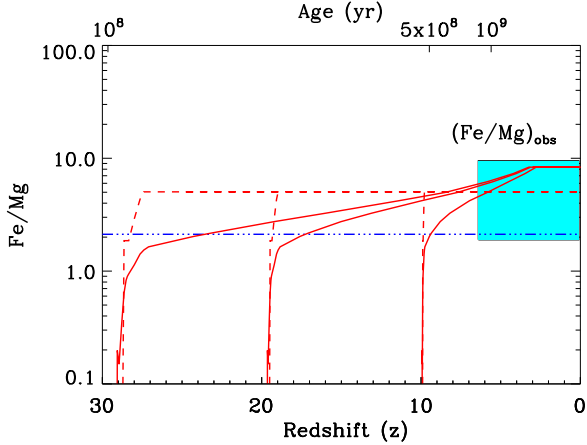,height=6.0cm}}}
\caption{Evolution of Fe/Mg as a function of redshift and
cosmological age. Dashed, solid and dashed-dotted lines represent
models A, B and C respectively (see Table 1). The three curves in
models A and B represent burst turn-on redshifts of $z$ = 30, 20 and
10. The shaded box indicates the approximate observed range of BLR gas
metallicities of 1--5 $Z_\odot$ from FeII/MgII over $z \sim$
0.1--6.4.}
\end{figure}

\begin{figure}
\center{{\epsfig{figure=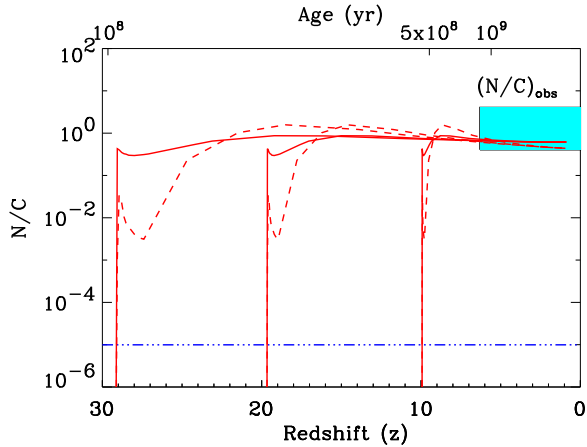,height=6.0cm}}}
\caption{Same as Figure 2, but for N/C.  The shaded box indicates the
  approximate observed range of BLR gas metallicities of 1--10 $Z_\odot$
  from NV/CIV over $z \sim$ 0.0--6.3.}
\end{figure}

%%%%%%%%%%%%%%%%%%%%%%%%%%%%%%%%%%%%%%%%%%%%%%%%%%%%%%%%%%%%%%%%%%%%%%%%%%

Our results may be summarized as follows: (i) model C (a top-heavy IMF with
PISNe) can explain the observed Fe/Mg values but not N/C, due to the severe
underproduction of N by PISNe. Note that the latter conclusion may change
with future revisions of VMS models; as Heger \& Woosley (2002) point out,
their exclusive study of helium cores may not adequately track N production
by VMSs.  (ii) Models A and B, which represent a present-day IMF with
nucleosynthetic contributions from Type II and/or Type Ia SNe, can
reproduce the observed ranges of both Fe/Mg and N/C at near- or supersolar
values, regardless of stellar metallicity. (iii) The inclusion of Type Ia
SNe pushes the predicted Fe/Mg towards the upper end of the observed values
(each Type Ia SN event generates Fe/Mg $\sim$ 87 and N/C $\sim$ 2.4 $\times
10^{-5}$), but is still within the observed range of Fe/Mg, as shown by
some earlier papers (see section 1). (iv) Even if alternate Type Ia SN
mechanisms are considered that lengthen the creation timescale beyond 1 Gyr
(see section 2), models A and B would still successfully reproduce the data
by $z \sim 6$, i.e., Type Ia SNe are not strongly required by the data.

These conclusions should be fairly robust since we have considered the
evolution of abundance ratios, which are independent of parameters like the
BLR mass, star formation efficiency, and the relative escape fractions of
metals from the BLR versus fallback into the central black hole.

\section{Discussion}

The study of element evolution in the deep potential wells associated with
the BLRs near the most massive detected high-$z$ objects, rather than in
the IGM (e.g., Ellison et al. 2000), has the potential advantage of not
being subject to the uncertainties of metal transport mechanisms. BLR
metals may therefore reflect the original element synthesis more
accurately, barring any enriched gas that falls into the central black
hole. Both IGM and BLR studies share, however, sizeable modelling
uncertainties in translating observed line strengths to the underlying
metallicity, which depends on the gas temperature, density and ionization
balance.

Within these uncertainties, our results indicate that stars in a
present-day IMF, regardless of stellar metallicity, can reproduce the data
on those QSO BLR abundance ratios that we have considered here. A new
population of stars (top-heavy IMF or otherwise) is not required from such
data, although VMSs at $z \sim$ 10--20 may be found to be attractive
candidates to explain other observations, e.g., in order to match the
observed amplitude and anisotropy of the near-infrared background
(Salvaterra \& Ferrara 2003; Magliocchetti, Salvaterra \& Ferrara 2003;
Cooray et al. 2003) and/or if the high Thomson optical depth initially
detected by $WMAP$ is not revised by future data.

Therefore, the epochs at $z \ga 6$ may not have been exclusively dominated
by a top-heavy IMF and in fact appears to require the presence of stars in
a present-day IMF in addition to any other coeval stellar population.
Indeed, a prompt transition from a top-heavy IMF to a Salpeter IMF may well
have occurred in the dense environments probed by QSO BLRs, as the metals
released by the first episodes of star formation could remain confined and
efficiently pollute the gas up to $Z_{\rm cr}$.  Not requiring an exclusive
VMS contribution by $z \sim$ 6 in QSO BLRs would also be consistent with,
e.g., the mild evolution of Fe/Mg over $z \sim$ 0.1--6.4, since metal-free
star formation is not observed at these lower redshifts.

If Type Ia SNe are assumed to be the dominant sources of iron, their
generation timescales in relation to the BLR Fe abundances at $z \sim$ 6
places the first star formation epochs at $z \sim$ 10--20 (Freudling et
al. 2003), consistent with $WMAP$'s findings. However, it is unclear if
Type Ia SNe can occur when $Z_\star=0$ stars are involved, and
consequently, barring extremely short metal enrichment timescales, whether
such SNe play a significant role in early cosmic metal generation.
Observations of large amounts of dust in the highest-$z$ QSOs (Bertoldi et
al. 2003) provide another avenue to constrain early SNe.  Although the
connection between the synthesis timescales of metals and dust is not yet
completely understood, it is currently believed that dust could be created
from Type II SNe (Todini \& Ferrara 2001) and/or PISNe (Schneider, Ferrara
\& Salvaterra 2003; Nozawa et al. 2003), but not from Type Ia SNe. As
Bertoldi et al. (2003) additionally point out, there may not be enough time
to create dust from intermediate-mass stars by $z \sim$ 6.

Finally, the ashes of the earliest generations of stars may be
reflected in the composition of very metal-poor halo stars in the
local universe. Observations of such low-mass stellar relics in our
Galaxy reveal the presence of elements heavier than Mg which cannot be
synthesized in their interiors, and which must therefore have been
already present in the pre-enriched gas cloud from which such stars
formed.  The recent detection of the most iron-poor star observed to
date (Christlieb et al. 2003) lies tantalizingly close to the
hypothesized $Z_{\rm cr}$ for the formation of VMSs (see section 2),
although this star is relatively enriched in other metals such as C.
This star's elemental abundance pattern indicates that the original
starforming cloud was enriched with metals ejected by SNe from $\sim$
20--130 $M_\odot$ stars (Umeda \& Nomoto 2003, Tumlinson, Venkatesan
\& Shull 2004), and it is only partially consistent with enrichment by
PISNe (Schneider et al. 2003a).  This derived mass range for a
first-generation stellar population is roughly similar to that which
could be responsible for the enrichment of QSO BLRs as shown in this
work. Thus, a top-heavy IMF may not be required to explain the
metal abundance patterns in QSO BLRs, which is consistent with the
current data on extremely metal-poor stars in the galactic halo. It
remains to be seen whether future observations of such local stellar
relics and of the highest-$z$ objects yield a consistent picture of
the sources of the earliest metal synthesis in the universe.

\section*{Acknowledgements}

We thank Fred Hamann, Mark Giroux and Roberto Maiolino for useful
correspondence. A.~V. is supported by an NSF Astronomy and Astrophysics
Postdoctoral Fellowship under award AST-0201670.

%%%%%%%%%%%%%%%%%%%%%%%%%
\begin{table}
\begin{center}
\caption{Total ejected mass in $M_{\odot}$ for C, N, Mg and Fe as a function of stellar mass and metallicity. 
For each element, we include the total abundances of all the relevant isotopes and, for Fe, the radioactive decay 
of $^{56}$Ni and $^{57}$Ni (see text).}
\begin{tabular}{|c|cccc|c|}\hline 
M   & $Z_\star$ &            C          &     N         	&       Mg         	&       Fe          	\\ \hline

1   &   0       & 1.78$\, 10^{-3}$	&  5.89$\, 10^{-5}$	&       0		& 	0		\\
    &   1	& 2.33$\, 10^{-3}$	&  1.52$\, 10^{-3}$	&  	0		&	0		\\	
1.25&   0       & 8.99$\, 10^{-3}$	&  8.87$\, 10^{-5}$	&       0		& 	0		\\
    &   1	& 2.71$\, 10^{-3}$	&  2.09$\, 10^{-3}$	&	0		&	0		\\
1.5 &   0       & 1.44$\, 10^{-2}$	&  1.18$\, 10^{-4}$	&       0		& 	0		\\
    &	1	& 5.69$\, 10^{-3}$	&  2.84$\, 10^{-3}$	&	0		&	0		\\
1.7 &   0       & 1.86$\, 10^{-2}$	&  1.41$\, 10^{-4}$	&       0		& 	0		\\
    &	1	& 7.44$\, 10^{-3}$	&  3.33$\, 10^{-3}$	&	0		&	0		\\
2   &   0       & 2.71$\, 10^{-2}$	&  2.32$\, 10^{-4}$	&       0		& 	0		\\
    &	1	& 1.12$\, 10^{-2}$	&  4.5$\, 10^{-3}$	&	0		&	0		\\
2.5 &   0       & 3.73$\, 10^{-2}$	&  3.12$\, 10^{-4}$	&       0		& 	0		\\
    &	1	& 1.88$\, 10^{-2}$	&  6.51$\, 10^{-3}$	&	0		&	0		\\
3   &   0       & 4.24$\, 10^{-2}$	&  3.57$\, 10^{-4}$	&       0		& 	0		\\
    &	1	& 2.63$\, 10^{-2}$	&  8.52$\, 10^{-3}$	&	0		&	0		\\
4   &   0       & 1.25$\, 10^{-2}$	&  5.66$\, 10^{-2}$	&	0		&	0		\\      
    &	1	& 3.33$\, 10^{-3}$	&  1.18$\, 10^{-2}$	&	0		&	0		\\
5   &   0       & 1.04$\, 10^{-2}$	&  9.28$\, 10^{-2}$	&	0		&	0		\\
    &	1	& 1.55$\, 10^{-3}$	&  4.22$\, 10^{-2}$	&	0		&	0		\\
7   &   0	& 2.40$\, 10^{-3}$	&  2.29$\, 10^{-2}$	&	0		&	0		\\
    &	1	& 2.26$\, 10^{-3}$	&  7.66$\, 10^{-2}$	&	0		&	0		\\
8   &	1	& 2.51$\, 10^{-2}$	&  9.60$\, 10^{-2}$	&	0		&	0		\\
11  &   1       & 5.42$\, 10^{-2}$	&  3.68$\, 10^{-2}$	&  1.22$\, 10^{-2}$	& 8.74$\, 10^{-2}$	\\ 
12  &  	0       & 4.3 $\, 10^{-2}$	&  5.54$\, 10^{-6}$	&  3.88$\, 10^{-3}$	& 8.55$\, 10^{-2}$	\\ 	
    &	1       & 8.15$\, 10^{-2}$	&  3.61$\, 10^{-2}$	&  1.11$\, 10^{-2}$	& 6.22$\, 10^{-2}$	\\
13  &	0       & 6.84$\, 10^{-2}$	&  1.01$\, 10^{-5}$	&  1.51$\, 10^{-2}$	& 0.20			\\
    &   1       & 0.12			&  4.68$\, 10^{-2}$	&  2.29$\, 10^{-2}$	& 0.16			\\
15  &	0       & 0.15			&  2.49$\, 10^{-5}$	&  3.41$\, 10^{-2}$	& 0.17			\\
    &	1       & 0.16			&  5.41$\, 10^{-2}$	&  3.99$\, 10^{-2}$	& 0.14			\\
18  &	0       & 0.11			&  2.29$\, 10^{-6}$	&  4.8$\, 10^{-6}$	& 5.47$\, 10^{-15}$	\\
    &	1       & 0.25			&  5.70$\, 10^{-2}$	&  7.71$\, 10^{-2}$	& 9.63$\, 10^{-2}$	\\
19  &   1       & 0.29			&  5.72$\, 10^{-2}$	&  4.59$\, 10^{-2}$	& 0.14			\\
20  &	0       & 8.98$\, 10^{-2}$	&  1.88$\, 10^{-6}$	&  2.04$\, 10^{-6}$	& 1.15$\, 10^{-17}$	\\
    &	1       & 0.21			&  6.0$\, 10^{-2}$	&  4.95$\, 10^{-2}$	& 0.12			\\
22  &	0       & 0.28			&  8.47$\, 10^{-5}$	&  9.51$\, 10^{-2}$	& 0.18			\\
    &	1       & 0.24			&  6.75$\, 10^{-2}$	&  6.24$\, 10^{-2}$	& 0.25			\\
25  &	0       & 8.31$\, 10^{-2}$    &  2.39$\, 10^{-4}$	&  5.61$\, 10^{-8}$	& 2.09$\, 10^{-10}$	\\
    &	1       & 0.32                  &  7.95$\, 10^{-2}$	&  0.16			& 0.17			\\
30  &	0       & 0.11                  &  3.65$\, 10^{-3}$	&  5.9$\, 10^{-7}$	& 2.83$\, 10^{-9}$	\\
    &	1       & 0.29                  &  0.10			&  0.27			& 2.86$\, 10^{-2}$	\\
35  &	0       & 9.79$\, 10^{-10}$   &  2.13$\, 10^{-8}$	&  2.30$\, 10^{-13}$	& 6.56$\, 10^{-36}$	\\
    &	1       & 0.30                  &  0.13			&  0.19			& 3.15$\, 10^{-2}$	\\
40  &	0       & 5.89$\, 10^{-10}$   &  2.66$\, 10^{-8}$	&  3.33$\, 10^{-13}$	& 1.16$\, 10^{-35}$	\\
    &	1       & 0.31                  &  0.14			&  0.12			& 3.23$\, 10^{-2}$	\\
140 &	0    	& 6.89 			&  7.91$\, 10^{-5}$   &  1.53 		&  2.20$\, 10^{-13}$	\\
149 &   0    	& 4.54 			&  5.74$\, 10^{-5}$   &  3.02 		&  0.04                	\\
158 &   0    	& 4.32 			&  4.98$\, 10^{-5}$   &  3.49 		&  0.15                	\\
168 &	0    	& 4.33 			&  4.82$\, 10^{-5}$   &  3.67 		&  0.17			\\
177 &	0    	& 4.28 			&  4.44$\, 10^{-5}$   &  3.97 		&  0.46			\\
186 &	0    	& 4.21 			&  4.11$\, 10^{-5}$   &  4.24 		&  1.38			\\
195 &   0       & 4.13 			&  3.87$\, 10^{-5}$   &  4.38 		&  3.08 		\\
205 &	0    	& 4.01 			&  3.74$\, 10^{-5}$   &  4.41 		&  5.98			\\
214 &	0    	& 3.85 			&  3.57$\, 10^{-5}$   &  4.4  		&  9.76			\\
223 &	0    	& 3.74 			&  3.43$\, 10^{-5}$   &  4.31 		&  14.48		\\
232 &	0    	& 3.73 			&  3.12$\, 10^{-5}$   &  4.5  		&  19.36 		\\
242 &	0    	& 3.71 			&  2.67$\, 10^{-5}$   &  4.55 		&  25.07		\\
251 &	0    	& 3.61 			&  2.11$\, 10^{-5}$   &  4.42 		&  32.33		\\
260 &	0    	& 3.49 			&  1.77$\, 10^{-5}$   &  4.38 		&  40.43		\\
\hline 
\end{tabular}
\end{center}
\end{table}
%%%%%%%%%%%%%%%%%%%%%%%%%%%%%%%%%%%%%%%%%%%%%%%

\end{document}